\documentclass{PoS}
\newcommand{\nc}[1]{\newcommand{#1}}

\nc{\its}[1]{\itshape #1 \upshape}
\nc{\mc}[3]{\multicolumn{#1}{#2}{#3}}

\nc{\bc}{\begin{center}}
\nc{\ec}{\end{center}}
\nc{\ig}[1]{\bc \includegraphics{#1} \ec}

\nc{\bo}[1]{\mbox{\boldmath \( #1 \! \! \)  \unboldmath}}
\newcommand{\beqn} {\begin{equation}}
\newcommand{\eqn} {\end{equation}}
\nc{\be}{\begin{eqnarray}}
\nc{\ee}{\end{eqnarray}}
\nc{\bew}{\begin{eqnarray*}}
\nc{\eew}{\end{eqnarray*}}
\nc{\bs}{\begin{subeqnarray}}   
\nc{\es}{\end{subeqnarray}}     
\nc{\nnn}{\nonumber \\}
\nc{\f}[2]{\frac{#1}{#2}}
\nc{\td}[2]{\f{d #1}{d #2}}
\nc{\pd}[2]{\f{\partial #1}{\partial #2}}
\nc{\suli}{\sum\limits}
\nc{\proli}{\prod\limits}
\nc{\ili}{\int\limits}
\nc{\sr}[2]{\stackrel{#1}{#2}}
\nc{\dps}{\displaystyle}
\nc{\ket}[1]{\left| #1 \right>}
\nc{\bra}[1]{\left< #1 \right|}
\nc{\bracket}[2]{\left< #1 \right| \left. \! #2 \right>}
\nc{\norm}[1]{\left\| #1 \right\|}
\nc{\lndm}[1]{\pd{^{#1} \ln{\det{M}}}{\mu^{#1}}}
\nc{\pdmm}[1]{M^{-1} \pd{^{#1} M}{\mu^{#1}}}
\nc{\pdm}{M^{-1}\pd{M}{\mu}}
\nc{\trac}[1]{\mbox{Tr}\left(#1\right)}
\nc{\hm}{\hat{m}}
\nc{\hmu}{\hat{\mu}}

\def\lsim{\raise0.3ex\hbox{$<$\kern-0.75em\raise-1.1ex\hbox{$\sim$}}}
\def\gsim{\raise0.3ex\hbox{$>$\kern-0.75em\raise-1.1ex\hbox{$\sim$}}}

\title{On Equation of State at physical quark masses}

\ShortTitle{On Equation of State}

\author{\speaker{Peter Petreczky (for RBC-Bielefeld Collaboration) }
         \thanks{This work has been supported by contract DE-AC02-98CH10886
          with the U.S. Department of Energy. The numerical calculations have been performed
          using QCDOC superomputers of USQCD Collaboration and RIKEN-BNL Research Center as well as 
           the BlueGene/L
          at the New York Center for Computational Sciences (NYCCS).}\\
        Physics Department, Brookhaven National Laboratory, Upton NY 11973\\
        E-mail: \email{petreczk@bnl.gov}}

\abstract{
QCD equation of state is calculated in (2+1) flavor QCD at  temperatures corresponding 
to the transition region with the physical values of the light quark masses using the p4 staggered fermion
action on lattices with temporal extent $N_{\tau}=8$.
The results are compared with previous calculations performed at twice
larger values of the light quark masses as well as with results obtained
from the resonance gas model calculation. The deconfining and
chiral aspects of the QCD transition are also discussed.
}

\FullConference{The XXVII International Symposium on Lattice Field Theory\\
                 July 26-31, 2009\\
                 Peking University, Beijing, China}

\begin{document}

\section{Introduction}
\label{intro}
In the past five years a lot of progress has been achieved in calculating QCD Equation of State
(for recent reviews see \cite{detar_lat08,petreczky_sewm06,petreczky_qm09}). 
In the most recent calculations the Equation of State (EoS)has been evaluated
for 2+1 flavor QCD, {\it i. e.} in QCD with one strange quark and two light ($u,d$) quarks using
various improved staggered fermion actions \cite{stout,MILCeos,ourEoS,hotqcd_eos}. 
The most extensive
calculations of the EoS have been performed with p4 and asqtad staggered fermion
formulations on
lattices with temporal extent $N_{\tau}=4,~6$ \cite{MILCeos,ourEoS} and $8$ \cite{hotqcd_eos}.
These actions improve both the taste symmetry of the staggered fermions as well as the 
quark dispersion relations. The latter insures that thermodynamic observables are
${\cal O}(a^2)$ improved at high temperatures and thus have only a small cut-off 
dependence in this regime. The stout-link action, which has been used for the calculation
of the EoS on lattices with temporal extent $N_{\tau}=4,~6$ \cite{stout}, only
improves the taste symmetry of the staggered fermions and therefore has the same large
discretization errors at high temperatures as the standard staggered fermion formulation. 
On the other hand the stout action has better taste symmetry which is important at low
temperatures.

While at high temperatures the masses of the relevant degrees of freedom, quarks and gluons,
are small compared to the temperature scale, this is not the case at low temperatures
and in the transition region. One thus may expect that at these temperatures thermodynamic 
observables are more sensitive to the quark masses, which control the mass of the
light pseudo-scalars and eventually are responsible for the occurrence of a true
phase transition in the chiral limit. Calculations with p4 and asqtad actions have so-far
been performed using light quark masses ($m_l$) which are one tenth of the strange 
quark mass ($m_s$) and correspond to a lightest pseudo-scalar Goldstone 
mass
of $220$ and $260$ MeV respectively \cite{hotqcd_eos}.
The calculations with the stout-link action have been performed at the physical value of 
the light quark mass.

The purpose of this work is to investigate the quark mass dependence of the EoS by 
calculating it with the p4-action for physical values of the (degenerate) light quark 
masses. The calculational procedure used in this work
closely follows that used in our previous calculations at $\hm_l=0.1\hm_s$ \cite{ourEoS}.

\section{Calculations of the Equation of State}
\begin{figure}
\includegraphics[width=8cm]{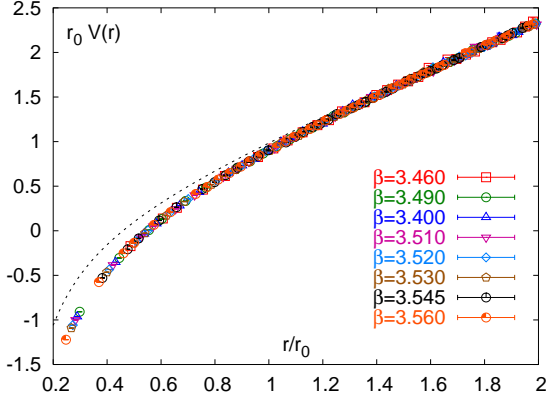}
\caption{The static potential in units of $r_0$ calculated at different gauge couplings.
The dashed line shows the string potential.}
\label{fig:v0}
\end{figure}

We have performed calculations with the p4-action for fourteen values of the gauge coupling
$\beta=6/g^2$ in the region of the finite temperature crossover. The finite temperature 
calculations have been performed  on $32^3\times 8$ lattices, while
the corresponding zero temperature calculations have been performed on $32^4$ lattices.
We used the physical value for the strange quark mass and the ratio of strange to 
light quark mass was chosen to be $h=m_s/m_l=20$. The lattice spacing was determined by
calculating the static potential and extracting  the Sommer scale $r_0$ from it.
To remove the additive 
divergent constant in the potential following Ref. \cite{ourEoS} we normalized it to the string potential 
$V_{string}(r)=-\pi/(12 r)+\sigma r$ at
distance $r=1.5r_0$. This is needed for the renormalization of the Polyakov loop as discussed in Ref. \cite{ourEoS}.
In Figure \ref{fig:v0} the static potential in units of $r_0$ and normalized to the string potential is shown. 
No discretization errors are visible in the potential.
We extracted pseudo-scalar meson masses 
using wall sources in the calculation of meson propagators.
It turned out that the $\eta_{s \bar s}$ mass
is, with 1 -2 \% accuracy, the same as in \cite{ourEoS}.
Thus, a re-adjustment of the line of constant physical $\eta_{s \bar s}$ mass 
corresponding to our new and smaller light quark masses was not
necessary.In fact,
in the present calculations we find that our quark mass values
define a line of constant physics characterized by
the following relations
$r_0 \cdot m_{\pi}=0.371(3)$, $r_0 \cdot m_K=1.158(5)$, $r_0 \cdot m_{\eta_{s\bar s}}=1.578(7)$
Using $r_0=0.469$ fm, as determined in Ref.\cite{gray}, we get $m_{\pi}=154$ MeV, 
$m_K=486$ MeV and\footnote{ A physical value for
the $\eta_{s \bar s}$ mass can be obtained from the relation 
$m_{\eta_{s\bar s}}=\sqrt{2 m_K^2-m_{\pi}^2} = 686$ MeV.}
$m_{\eta_{s\bar s}}=663$~MeV. 
This means that both the light quark masses
and the strange quark mass are very close to their physical values. Furthermore, 
in the entire parameter range covered by our thermodynamic calculations deviations of the 
meson masses from the above values are less than $3\%$.

The calculation of the EoS starts with the evaluation of
the trace anomaly, {\it i.e.} the trace of the energy-momentum tensor $\Theta_{\mu\mu}(T)$.
It is related to the temperature derivative of the pressure through thermodynamic identities, 
\begin{equation}
\frac{\Theta_{\mu\mu}(T)}{T^4}=\frac{\epsilon-3p}{T^4}=T\frac{{\rm d}}{{\rm d}T} \left( \frac{p}{T^4}\right)\; .
\label{e-3p}
\end{equation}
The trace anomaly can be expressed in terms of the expectation values of quark condensates
and the gluon action density, see e.g. Ref. \cite{hotqcd_eos}.
\
\begin{figure}
\includegraphics[width=8cm]{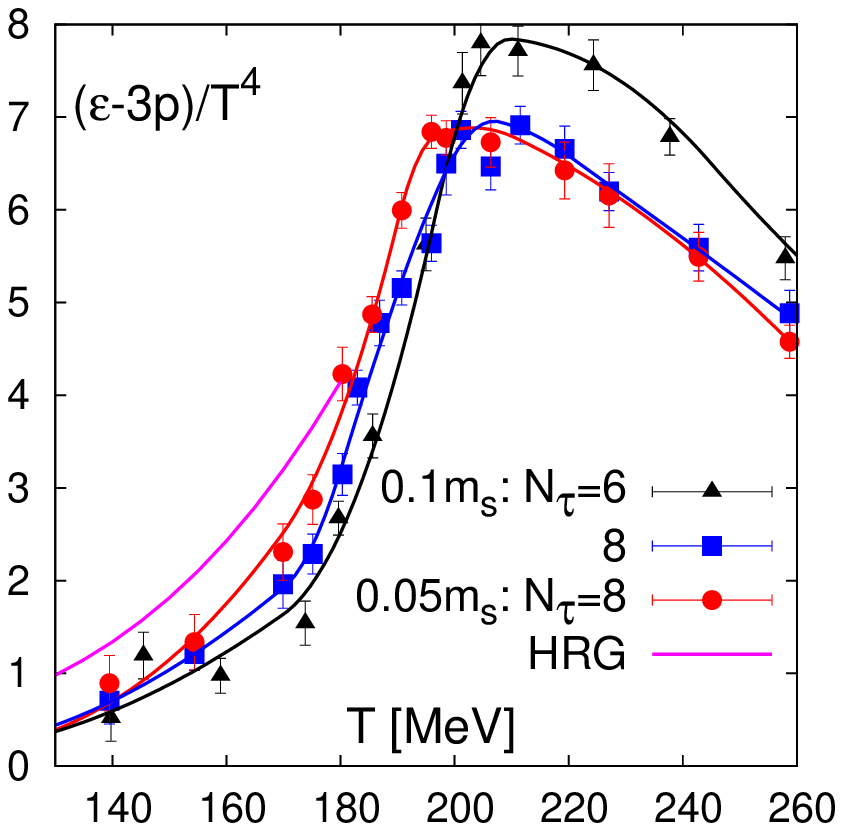}\hspace*{-1cm}
\includegraphics[width=8cm]{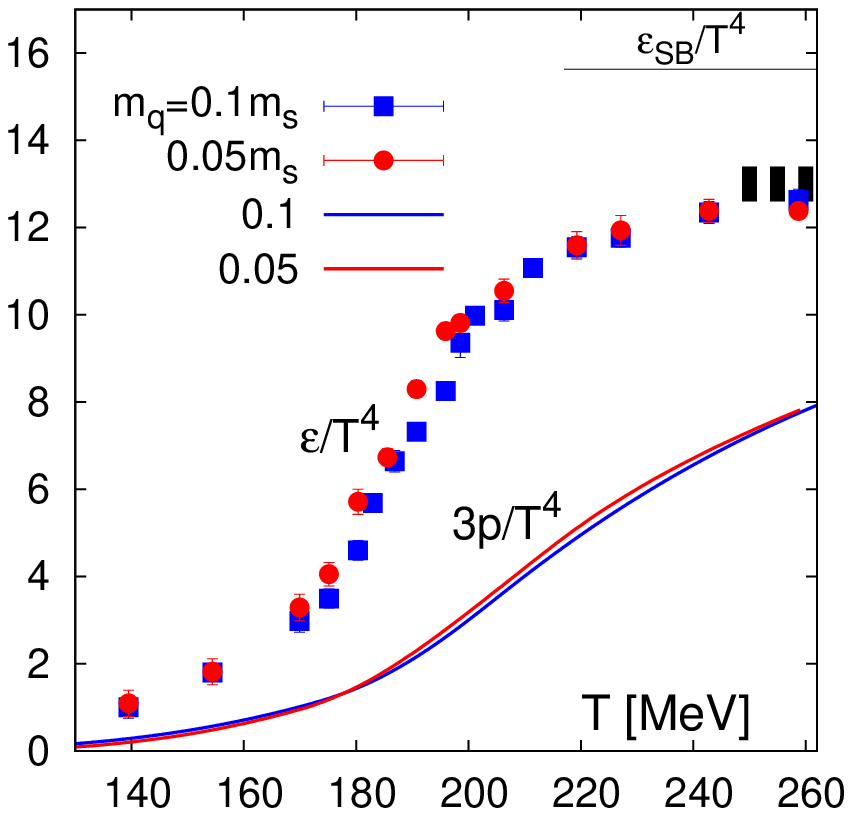}
\caption{The trace anomaly $(\epsilon-3p)/T^4$ calculated for
the physical quark mass and compared with previous calculations at larger
light quark masses $m_l=0.1m_s$ (left) and the pressure as well as the energy density (right).
We compare the trace anomaly with as well as with the HRG model, which
includes all the resonances up to $2.5$GeV.
The horizontal band shows the expected uncertainty in the energy
density due to the choice of the lower integration limit}
\label{fig:e-3p}
\end{figure}
The numerical results
are shown in Figure \ref{fig:e-3p} and are compared to the previous calculation at twice
larger quark mass $m_l=0.1m_s$ on $N_{\tau}=6$ lattices \cite{ourEoS} and $N_{\tau}=8$
lattices \cite{hotqcd_eos}. The differences between $N_{\tau}=6$ and $N_{\tau}=8$ calculations
are due to cutoff effects and have been discussed in Ref. \cite{hotqcd_eos}.As one can see from the figure 
the main differences to the $N_\tau=8$ results at $m_l = 0.1 m_s$
arise for temperatures $T < 200$ MeV.
These differences can be understood 
as resulting from an expected shift of the transition temperature
by 5 MeV when the light quark mass is lowered to approximately
its physical value. 
At lower temperatures
it also is expected that the trace anomaly increases
with decreasing quark masses as hadrons become lighter when the quark mass is decreased.
While a tendency for such an increase may be indicated by the data at the lowest two
temperatures reached in our calculation, this effect is 
certainly not significant within the current statistical accuracy.

At temperatures below the transition temperature it is expected that thermodynamic 
quantities are well described by a hadron resonance gas (HRG) model. In fact, the freeze-out of 
hadrons in heavy ion experiments takes place in the transition region and the observed 
particle abundances are well described by the HRG model \cite{cleymans,pbm}. 
Therefore in Figure \ref{fig:e-3p} we also show the prediction of the HRG model, which
includes all the known resonances up to the mass $M_{max}=2.5$ GeV. 
The lattice data for $\epsilon-3p$ are below
the HRG prediction although the deviations from it are smaller compared to the results obtained
at $m_l=0.1m_s$. 
We mention again the present statistical accuracy and the possibility of
discretization effects in the hadron spectrum. In particular, due to taste breaking of staggered fermions
pseudo-scalar mesons are not degenerate at finite lattice spacing, therefore their contribution to thermodynamic
quantities maybe suppressed.

From the trace anomaly the pressure and thus other thermodynamic quantities can be calculated 
by performing the integration over the temperature
\beqn
\frac{p(T)}{T^4}-\frac{p(T_0)}{T_0^4}=\int_{T_0}^{T} d T' \frac{1}{{T'}^5} \Theta_{\mu \mu}(T').
\label{p_int}
\eqn
Here $T_0$ is an arbitrary temperature value that is usually chosen in the 
low temperature regime where the pressure and other thermodynamical quantities 
are suppressed exponentially by Boltzmann factors associated with the lightest
hadronic states, i.e. the pions.
Energy $\epsilon$ and entropy ($s T = (p+ \epsilon)$) 
densities are then obtained by combining results for $p/T^4$ and $(\epsilon-3p)/T^4$.
The numerical results for the pressure and energy density are shown in Fig. \ref{fig:e-3p}.
The uncertainties from the choice of the lower integration limit are shown as a horizontal band
in the figure.  The estimated uncertainties are about $8\%$ in the energy density
at the highest temperature of $T \simeq 260$MeV, and about $13\%$ for the pressure.

\section{Deconfinement and chiral aspects of the QCD transition}
\label{sec:deconf}

In the previous section we have seen that the energy density shows a rapid
rise in the temperature interval $T=(170-200)$MeV. This is
usually interpreted to be due to deconfinement, i.e.
liberation of many new degrees of freedom. For sufficiently large quark mass
this transition is known to be a first order transition (see e.g. Ref. \cite{stickan}).
In the limit of infinitely large quark mass the
order parameter for the deconfinement
phase transition  is the Polyakov loop. After renormalization it can be related
to the free energy of a static quark anti-quark pair $F_{\infty}(T)$ at infinite separation 
\cite{mclerran81,okacz02}
$L_{ren}(T)=\exp(-F_{\infty}(T)/(2 T))$
A rapid change in this quantity   is indicative for deconfinement
also in the presence of light quarks. 
In the opposite limit of zero quark mass 
one expects a chiral transition
and the corresponding 
order parameter is the chiral condensate.
For a genuine phase transition, i.e. in the chiral limit
the chiral condensate vanishes at the critical temperature $T_c$. However, we expect
that even for the crossover at finite quark mass the light quark
condensate rapidly drops in the transition region, indicating an
approximate restoration of the chiral symmetry. At non-vanishing quark mass
the chiral condensate needs additive and multiplicative
renormalization. Therefore, following Ref. \cite{ourEoS,hotqcd_eos} we introduce the so-called subtracted
chiral condensate
\begin{equation}
\Delta_{l,s}(T)=\frac{\langle \bar\psi \psi \rangle_{l,\tau}-\frac{m_l}{m_s} \langle \bar \psi \psi \rangle_{s,\tau}}
{\langle \bar \psi \psi \rangle_{l,0}-\frac{m_l}{m_s} \langle \bar \psi \psi \rangle_{s,0}}.
\end{equation}
Here the subscripts $l$ and $s$ refer to light and strange chiral condensates, while the 
subscript $0$ and $\tau$ to the case of zero and finite temperature respectively.
Subtraction of the strange quark condensate multiplied by the ratio of the light to strange quark
mass removes the quadratic divergence proportional to the quark mass. 

\begin{figure}
\includegraphics[width=7cm]{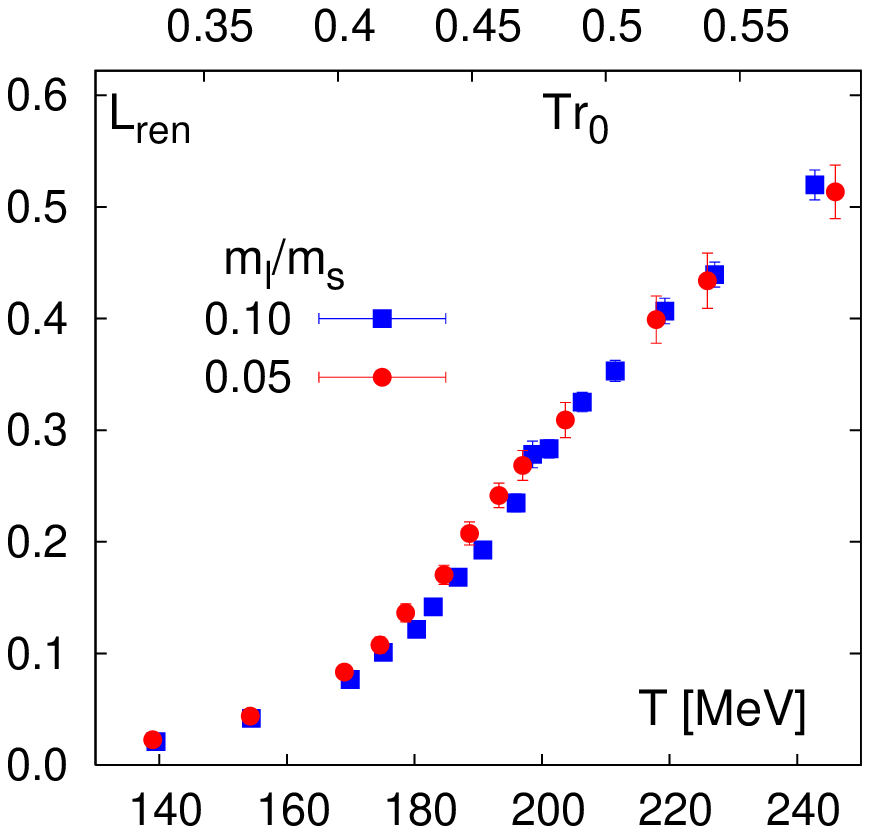}
\includegraphics[width=9.2cm]{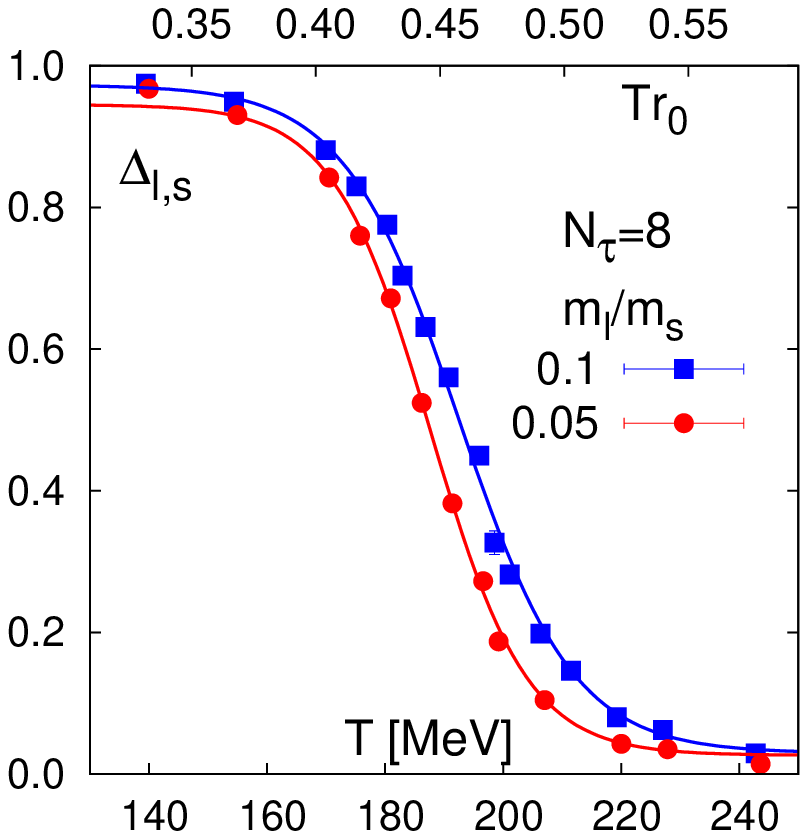}
\caption{The renormalized Polyakov loop (left) and the subtracted chiral condensate (right) as function of the
temperature calculated at $m_l=0.05m_s$ and at $0.1m_s$.}
\label{fig:orderpar}
\end{figure}

In Figure \ref{fig:orderpar} 
we show the renormalized Polyakov loop
and the subtracted chiral condensate $\Delta_{l,s}$ and compare with previous calculations performed at light
quark masses equal to one tenth of the strange quark mass \cite{hotqcd_eos}.
The renormalized Polyakov loop rises in the temperature interval $T=(170-200)$ MeV
where we also see the rapid increase of
the energy density. At the same time the subtracted chiral condensate rapidly drops in the 
transition region, indicating
that the approximate restoration of the chiral symmetry happens in the same temperature interval as deconfinement.
Compared to the calculation performed at light quark masses equal to one tenth of the strange quark mass we see
a shift of the transition region by roughly $5$ MeV. 
We note that such a shift arises differently in different observables.
In the case of the subtracted chiral condensate, for instance, a major ingredient to the
'shift' is the fact, that at fixed temperature
the condensate in the transition region is strongly quark mass dependent and drops
proportional to $\sqrt{m_l/m_s}$ \cite{Goldstone}.

\begin{figure}
\includegraphics[width=7cm]{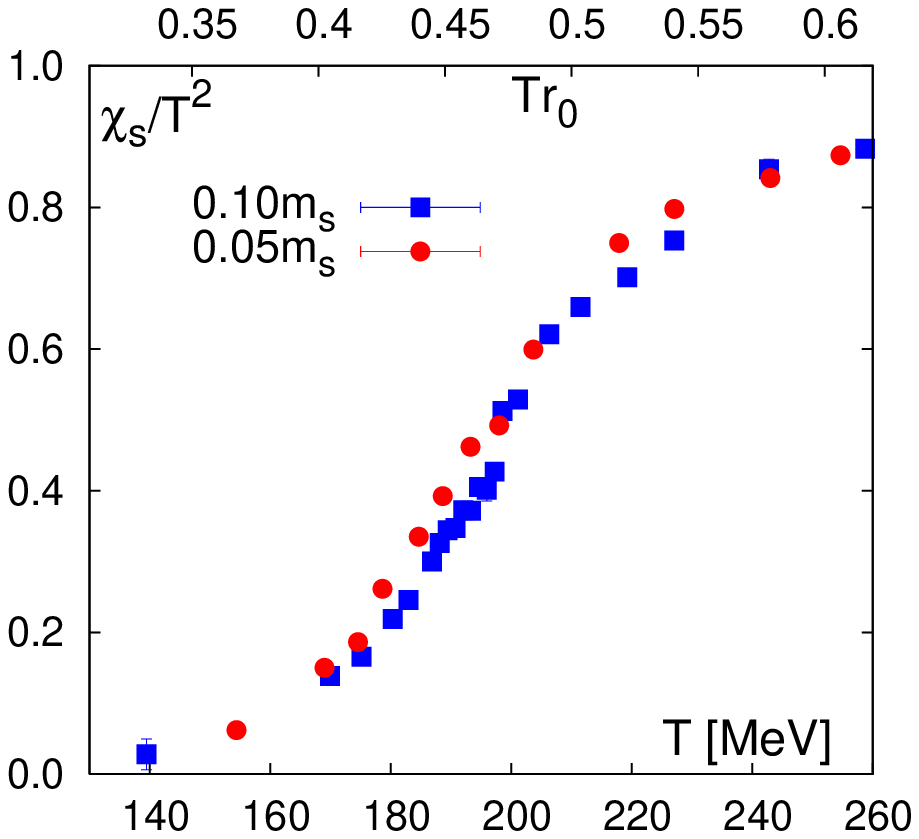}
\includegraphics[width=7cm]{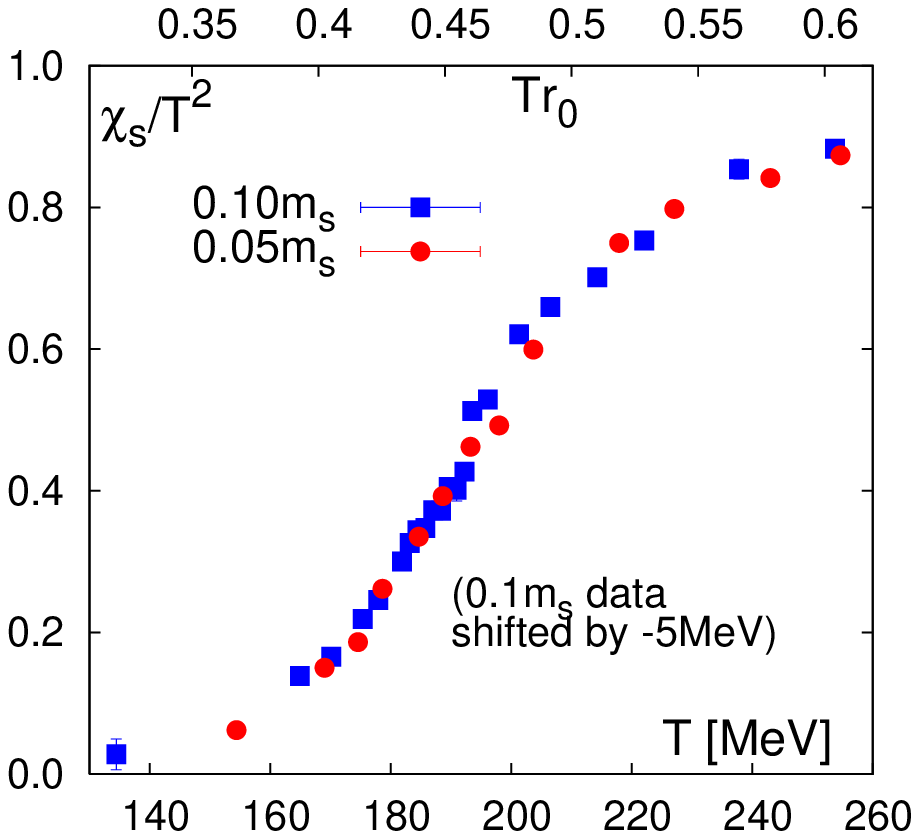}
\caption{Strangeness fluctuations as function of the
temperature calculated at $m_l=0.05m_s$ and at $0.1m_s$.
In the right plot the numerical data for $m_l=0.1m_s$ have been shifted
by $5$ MeV.
}
\label{fig:chis}
\end{figure}

The fluctuation of strangeness is also indicative of deconfinement. It can be defined
as the second derivative of the free energy density with respect to the strange quark
chemical potential
\begin{equation}
\chi_s(T)=\frac{1}{T^3 V}\frac{\partial^2 \ln Z(T,\mu_s)}{\partial \mu_s^2}|_{\mu_s=0}.
\end{equation}
At low temperatures 
strangeness is carried by massive hadrons and therefore strangeness fluctuations are
suppressed. At high temperatures strangeness is carried by quarks and 
the effect of the strange quark mass is small. Therefore strangeness fluctuations are not suppressed
at high temperatures. As discussed in Ref. \cite{hotqcd_eos} strangeness fluctuations behave
like the energy density in the transition region, i.e. they rapidly rise in a narrow temperature 
interval. In Fig. \ref{fig:chis} we show the strangeness fluctuations calculated at $m_l=0.05m_s$
and compare them with previous calculations performed at $m_l=0.1m_s$ \cite{hotqcd_eos}.
In the  bottom figure we also show the strangeness fluctuation for $m_l=0.1m_s$ with a
$5$ MeV shift of the temperature scale. As one can see this shift accounts for most of the 
quark mass dependence of the strangeness fluctuations. This is consistent with the conclusion
obtained from the quark mass dependence of other thermodynamic observables. 

\section{Conclusion}

We have calculated the EoS, renormalized Polyakov loop, subtracted chiral condensate and
strangeness fluctuations in (2+1)-flavor QCD in the crossover region from low to
high temperatures using the
improved p4 staggered fermion formulation on lattices with temporal extent $N_{\tau}=8$
at physical values of the light and strange quark masses. 
We found that thermodynamic quantities below the deconfinement 
transition are larger compared to the previous
calculations performed at twice larger quark mass but fall below the
resonance gas model result. 
The differences in the thermodynamic quantities calculated at
$m_l=0.05m_s$ and $m_l=0.1m_s$ can be well understood in terms 
of the shift of the transition temperatures
towards smaller values when the quark mass is decreased.
This conclusion is also supported by the calculation of renormalized
Polyakov loop, subtracted chiral condensate and strangeness fluctuations.
No additional enhancement of the pressure and the energy density is 
seen at low temperatures. 
This and the deviation from the resonance gas model
may be a cutoff effect due to taste violations. However, better 
statistical accuracy and calculations at smaller lattice spacing 
are needed to quantify this assertion. The transition region in our calculations
corresponds to larger temperatures compared to recent calculations with stout action
\cite{fodor06,fodor09}. It remains to be seen whether the taste symmetry violations which
are larger for the p4 action are responsible for this discrepancy.
At temperatures above $200$ MeV no quark mass dependence 
is seen in the equation of state.

\end{document}